# PLANS D’INFORMATION DE KULLBACK-LEIBLER MINIMALE

Astrid Jourdan
*Chercheur associé au LMA*
*Equipe de Probabilités et Statistiques*

# Introduction

Depuis quelques années, la simulation numérique modélise des phénomènes toujours plus complexes. De tels problèmes, généralement de très grande dimension, exigent des codes de simulation sophistiqués et très coûteux en temps de calcul (parfois plusieurs jours).

Dans ce contexte, le recours systématique au simulateur devient illusoire. L’approche actuellement privilégiée consiste à définir un nombre réduit de simulations organisées selon un plan d’expériences numériques et d’adapter, à partir de celui-ci, un métamodèle pour approcher le simulateur.

Dans le cadre de ces travaux, nous nous sommes intéressés à la construction des plans d’expériences en phase exploratoire i.e. lorsque la dépendance entre les entrées et les sorties est *a priori* inconnue. Il est alors difficile de prévoir quel type de métamodèle va convenir. Les plans élaborés lors de cette phase exploratoire doivent donc s’affranchir de toute contrainte par rapport à un type de métamodèle (régression linéaire, krigeage, réseaux de neurones) puisque celui-ci est choisi à l’issue de ces premières simulations.

Ainsi, les expériences de ces plans doivent remplir au mieux l’espace des paramètres afin d’obtenir des informations dans tout le domaine expérimental et notamment pour détecter les éventuelles irrégularités à l’intérieur du domaine de simulation. On cherche ainsi un plan dont les points seraient le plus uniformément répartis dans l’hypercube unité. Similairement à la discrépance, l’information de Kullback-Leibler (information KL) permet de mesurer l’écart entre la distribution empirique et la loi uniforme. L’idée est alors de construire, de façon empirique, des plans d’information KL minimale à l’aide d’un simple algorithme d’échange.

Ce rapport est organisé comme suit. Dans un premier temps, nous présentons l’information de Kullback-Leibler ainsi que les propriétés qui lui sont associées, notamment son application dans le cas de la loi uniforme et son lien avec l’entropie. Nous proposons ensuite deux méthodes de construction de plans d’information KL minimale basées sur des méthodes d’estimation de l’entropie différentes. La première utilise une estimation par noyaux de la fonction de densité afin de l’intégrer dans un calcul Monte Carlo. La deuxième évite l’estimation de la fonction de densité en passant par une méthode de plus proche voisin. Enfin, nous terminons par une étude comparative avec les plans usuels à l’aide de critères couramment utilisés dans la littérature.

# 1. Information de Kullback-Leibler

## 1.2. Information de Kullback-Leibler

Supposons que les points du plan sont les *n* réalisations d'un vecteur aléatoire $X=(X_1,\ldots,X_d)$ de fonction de densité inconnue *f* de support le cube unité $E=[0,1]^d$. L'objectif est de construire un plan dont la fonction de densité associée est la plus proche possible de celle de la loi uniforme sur E. Pour cela, nous avons choisi d'utiliser l'information de Kullback-Leibler (1951) afin de mesurer l'écart entre les fonctions de densité.

De façon générale, l'information de Kullback-Leibler permettant de mesurer l'écart entre les deux fonctions de densité *f* et *g* est définie par

$$\mathrm{I}(f,g)=\int f(\mathrm{x})\ln\left(\frac{f(\mathrm{x})}{g(\mathrm{x})}\right)\mathrm{dx}$$

L'intégrale ci-dessus n'est pas toujours définie. Une condition nécessaire pour que l'intégrale converge est que $P_f$, la mesure de probabilité sous-jacente à la fonction de densité *f*, est absolument continue par rapport à $P_g$ la mesure de probabilité induite par *g*. Dans notre cas, si on considère *f* continue et *g* la fonction de densité de la loi uniforme sur E, cela revient à supposer que pour tout $A\subset\mathbb{R}^d\backslash E$, $P_f(A)=0$.

## 1.1. Quelques propriétés de l'information de Kullback-Leibler

Nous présentons ici les propriétés de l'information de Kullback-Leibler nécessaires à la bonne compréhension de nos propos.

**<u>Propriétés de distance</u>**

Même si l'information de Kullback-Leibler est utilisée pour mesurer un écart entre deux fonctions de densité, elle n'est pas une distance au sens topologique du terme. En effet, l'inégalité triangulaire n'est pas vérifiée ainsi que la propriété de symétrie. Ceci explique par ailleurs l'importance du choix entre les fonctions *f* et *g*. Par convention, *f* est choisie comme la fonction de densité associée au plan et *g* comme la fonction de densité théorique, à savoir ici celle de la loi uniforme. Néanmoins, il existe des modifications pour symétriser l'information KL, par exemple en utilisant,

$$\mathrm{d}(f,g)=\mathrm{I}(f,g)+\mathrm{I}(g,f),$$

on parle alors de divergence de Kullback-Leibler.

En revanche, l'information de Kullback-Leibler est définie positive

$$\mathrm{I}(f,g)\geq 0$$

avec égalité si et seulement si $f=g$ presque partout. Cette propriété justifie en partie l'utilisation de l'information de Kullback-Leibler pour mesurer l'écart entre deux fonctions de densité. Ainsi, plus l'information est proche de 0, et plus *f* est « proche » de *g*.

**<u>Formulation sous forme d'espérance</u>**

En développant l'intégrale, on obtient

$$I(f,g)=\int f(x)\ln(f(x))dx-\int f(x)\ln(g(x))dx\,.$$

Ce qui peut encore s'exprimer sous la forme d'une espérance relativement à la mesure $P_f$,

$$I(f,g)=E_{P_f}\left[\ln(f(x))\right]-E_{P_f}\left[\ln(g(x))\right]$$

**Propriété d'additivité**

L'information K-L est additive, c'est-à-dire que l'information entre deux lois conjointes de variables i.i.d. est la somme de l'information des lois marginales. Si on suppose que les coordonnées des points du plan sont variables i.i.d. alors la propriété d'additivité de l'information KL revient à tester la distribution uniforme des points du plan sur les axes uniquement. Or une répartition uniforme sur les marges n'entraîne pas nécessairement une répartition uniforme dans le cube unité.

**Changement d'échelle**

On note de plus que l'information KL est invariante par changement d'échelle. Cette propriété permet de travailler dans le cube unité pour construire des plans d'expériences génériques.

## 1.2. Maximisation de l'entropie

Le lien entre l'information de Kullback-Leibler et l'entropie se fait immédiatement en appliquant la définition à notre cas, c'est-à-dire en considérant $g$ comme la fonction de densité de la loi uniforme sur E,

$$I(f)=\int f(x)\ln(f(x))dx=E_{P_f}\left[\ln(f(x))\right]=-H[f]$$

où H[$f$] est l'entropie. Ainsi, minimiser l'information KL revient à maximiser l'entropie.

On retrouve la définition des plans à entropie maximale couramment utilisés en planification numérique (Shewry & Wynn, 1987, Currin *et al.*, 1988) ou plans bayésiens optimaux (Chaloner & Verdinelli ,1995, Sebastiani & Wynn , 2000).

L'originalité de l'idée proposée ici vient du fait que nous sommes en phase exploratoire et que nous n'avons aucun modèle sous-jacent. La maximisation de l'entropie n'a donc pas pour objectif d'augmenter la quantité d'information (au sens de Shannon) contenue dans l'échantillon relativement à des paramètres du modèle. En revanche, il est bien connu que la loi uniforme maximise l'entropie des lois à support dans [0,1]. L'entropie du plan est donc négative et faire tendre cette entropie vers 0, revient à s'approcher d'une distribution uniforme. On notera d'ailleurs que des tests d'uniformité ont été développés à partir de l'estimation de l'entropie (Dudewicz *et al.* 1981, voir autres réf) mais uniquement dans le cas pour la dimension un et ne peuvent se généraliser en dimension supérieure. Afin de ne pas confondre avec les plans à entropie maximale, les plans construits dans ce rapport seront dits d'information KL minimale.

<u>Remarque</u> : On note, dans ce cas, que la divergence de Kullback-Leibler s'écrit

$$d(f)=\int f(x)\ln(f(x))dx-\int f(x)dx=E_{P_f}\left[\ln(f(x))-1\right]$$

### 1.3. Construction des plans

Au vu des propriétés énoncées précédemment, les plans sont construits suivant un simple algorithme d'échange visant à maximiser l'entropie.

**Algorithme d'échange**

**Initialisation :** choisir un plan d'expériences $x_1,\ldots,x_n$ selon une loi de probabilité donnée

**TANT QUE (condition d'arrêt n'est pas satisfaite)**

Choisir une expérience $x_i$ au hasard parmi les points du plan

Remplacer $x_i$ par $y_i$ simuler uniformément dans $[0,1]^d$

**SI** l'entropie obtenue à l'action précédente est plus grande que celle du plan d'origine

**ALORS** accepter le changement $x_i=y_i$ .

**FIN TANT QUE**

La condition d'arrêt n'est pas spécifiée. Pratiquement, on considère soit un nombre maximum d'échanges, soit la stabilisation de l'entropie qui se traduit par un nombre arbitraire d'échanges consécutifs testés et refusés. Suite à l'étude de la convergence de l'algorithme sur plusieurs essais, le nombre maximal d'échanges est fixé à $1000\times d$ et le nombre d'échanges consécutifs sans amélioration à $100\times d$. Le plan final dépend plus ou moins du plan initial choisi aléatoirement. Afin de réduire cette dépendance, plusieurs initialisations sont testées et le meilleur plan est sélectionné. Dans les paragraphes suivants, plusieurs plans sont construits à partir d'une seule initialisation de façon à étudier l'influence du plan initial.

Il est évident que dans l'algorithme ci-dessus le point essentiel concernant le calcul de l'entropie d'un ensemble de points reste à définir. Il existe différentes techniques d'estimation de l'entropie qui, pour certaines d'entre elles, sont exposées dans l'article de Beirlant *et al.* (1997). Nous avons retenu deux méthodes qui permettront de construire deux types de plans.

# 2. Estimation de l'entropie par Monte Carlo

### 2.2. La méthode de Monte Carlo

Notons $X_1,\ldots,X_n$, les *n* réalisations de X constituant les points du plan. Etant donné que l'entropie s'écrit sous la forme d'une espérance,

$$\mathrm{H}(\mathrm{X}) = -\mathrm{E}\left[\ln\left(f(\mathrm{X})\right)\right],$$

la méthode de Monte Carlo fournit un estimateur sans biais et convergent de l'entropie

$$\hat{\mathrm{H}}(\mathrm{X}) = -\frac{1}{n}\sum_{i=1}^{n}\ln f(\mathrm{X}_i).$$

Cette estimation fait intervenir la fonction de densité *f* inconnue mais pouvant être estimée à partir de $X_1,\ldots,X_n$. La solution consistant à remplacer *f* par une estimation dans l'expression ci-dessus, n'assure pas que l'estimateur reste sans biais.

Nous avons choisi ici d'estimer la fonction de densité par une méthode à noyaux (Silverman 1986, Scott 1992),

$$\hat{f}(x) = \frac{1}{nh^d}\sum_{i=1}^{n}\mathcal{K}\left(\frac{x - X_i}{h}\right),\ \forall x \in [0,1]^d$$

où *h* est la taille de la fenêtre et $\mathcal{K}$ est le noyau.

Remarque : L'estimation de la fonction de densité et l'estimation de l'entropie se font à partir du même échantillon, *i.e* les points du plan. Dans le contexte des plans d'expériences, l'échantillon est de petite taille. Or, il n'y a pas d'exigence à ce que l'estimation de l'entropie se fasse à partir des points du plan. On pourrait envisager la méthode de Monte Carlo suivante (Dmitriev, Tarasenko,1973)

$$\hat{H}(X) = -\frac{1}{N}\sum_{i=1}^{N}\hat{f}(Z_i)\ln\hat{f}(Z_i),$$

où $Z_i$ sont N réalisations d'un vecteur aléatoire de loi uniforme sur $[0,1]^d$. Dans ce cas, seule l'estimation de la fonction de densité dépend des points du plan. Il est alors possible de choisir un échantillon de grande taille N pour l'estimation de l'entropie. Cette méthode n'a pas été retenue car, comme le souligne Joe (1989), elle nécessite un temps de calcul prohibitif lorsque $d \geq 2$. De plus les résultats de tests effectués en dimension un ne s'avèrent pas meilleurs.

## 2.3. Taille de la fenêtre

La taille de la fenêtre a une grande influence sur la qualité de l'estimation. En supposant que les variables sont non corrélées, il est assez usuel d'estimer la taille de la fenêtre par (Règle de Scott, 1992)

$$\hat{h}_j = n^{-1/(d+4)}\hat{\sigma}_j,\ j=1,\ldots,d$$

où $\hat{\sigma}_j$ est l'écart-type de la jème variable aléatoire $X_j$.

Cependant, Joe (1989) montre que dans le cas où f est estimée par noyaux, l'estimateur de l'entropie par la méthode de Monte Carlo est biaisé. Le biais dépend naturellement de la taille de l'échantillon *n*, de la dimension *d*, mais aussi de la taille de la fenêtre *h*. Lors de la construction d'un plan optimal, la taille *n* et la dimension *d* sont fixées. Il reste donc à fixer la taille de la fenêtre de façon à ce que le biais ne varie pas au cours de l'algorithme d'échange.

Pour cela, nous remplaçons $\hat{\sigma}_j$ par l'écart-type de la loi uniforme sur [0,1], d'où

$$\hat{h} = \frac{1}{\sqrt{12}}\frac{1}{n^{1/(d+4)}}.$$

## 2.4. Choix du noyau

Il est connu que la forme du noyau a peu d'influence sur l'estimation. Cependant pour des raisons liées à notre contexte et que nous évoquons ci-dessous, différents noyaux ont été testés.

Noyau d'Epanechnikov

Le noyau d'Epanechnikov multidimensionnel est défini par

$$\mathcal{K}(z) = \begin{cases} \alpha\left(1 - \|z\|^2\right) & \text{si } \|z\| \leq 1 \\ 0 & \text{sinon} \end{cases}$$

avec $\alpha$ la constante de normalisation (annexe 2) et $\|.\|$ la norme euclidienne.

<u>Noyau gaussien</u>

Le noyau gaussien centré multidimensionnel est défini par

$$\mathcal{K}(z) = \frac{(2\pi)^{-d/2}}{s^d} \exp\left[-\frac{1}{2s^2}\|z\|^2\right],$$

où $s^2$ est choisi égal à la dimension multipliée par la variance de la loi uniforme sur [0,1],

$$s^2 = \frac{d}{12}.$$

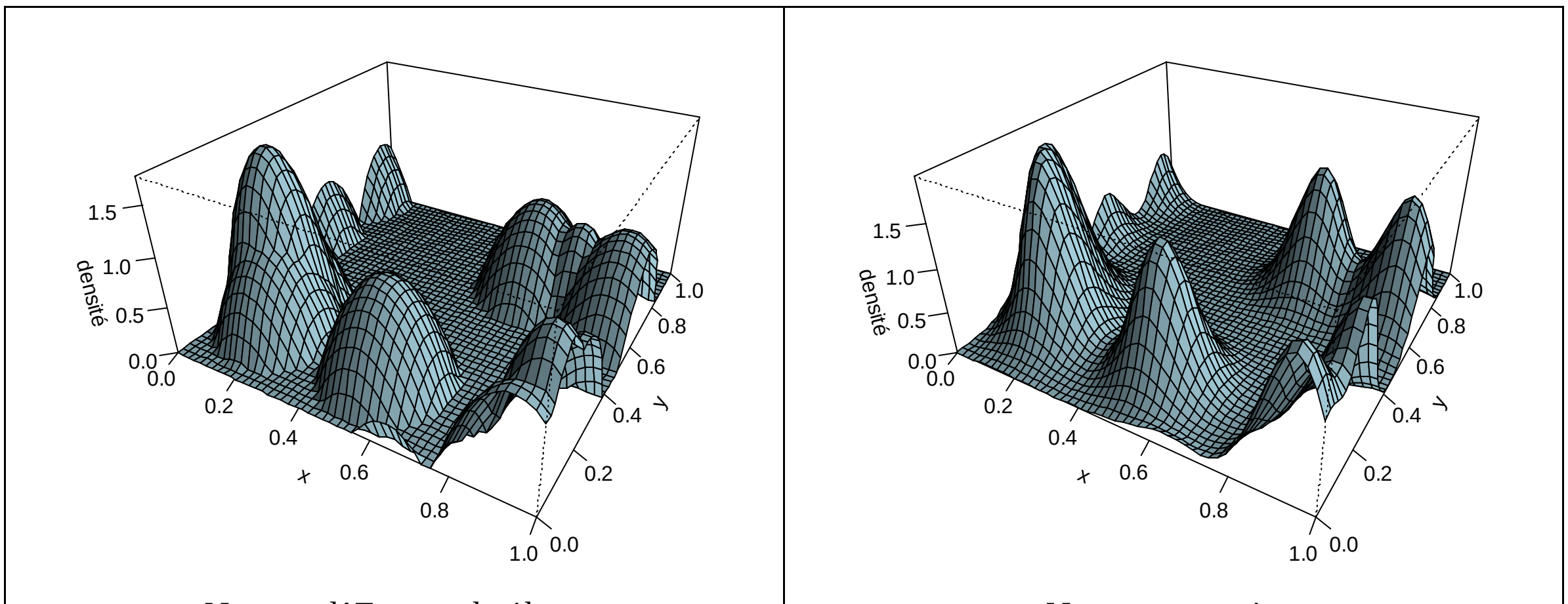

**Fig. 1. Représentation d'une fonction de densité estimée sur un échantillon de taille 20 en dimension 2.**

On constate sur la figure 1 que la forme du noyau n'a effectivement que peu d'influence sur le résultat. Le noyau d'Epanechnikov est à support borné et convient ainsi à notre problématique. On note cependant que cette caractéristique pose problème en dimension élevée car la probabilité que la norme de *z* soit inférieure à 1 devient très faible. En effet, si on suppose que la taille de la fenêtre est fixée comme expliqué ci-dessus alors

$$P(\|z\| < 1) = \frac{1}{12d} \frac{1}{n^{2/(d+4)}}.$$

Par exemple pour $d$=3 et $n$=30, cette probabilité est de $10^{-1}$, et pour $d$=10 et $n$=100, elle vaut $4\times10^{-3}$ (annexe 1). Cela signifie que le noyau est toujours nul et ainsi la fonction de densité estimée est identique quel que soit l'échantillon utilisé, ce qui n'apporte pas d'intérêt à l'utilisation de l'algorithme d'échange. Etant à support infini, le noyau gaussien ne présente pas ce type de problème et est finalement retenu pour l'estimation de la fonction de densité.

Les propriétés établies par Joe (1989) concernant cet estimateur sont asymptotiques. De plus, il montre que la taille de l'échantillon augmente rapidement avec la dimension pour avoir une estimation correcte de l'entropie (cf.

figure 2). Dans le contexte des plans d'expériences, le nombre de points est faible et donc le calcul de l'entropie sera entaché d'une erreur importante. Cependant, notre objectif n'est pas d'obtenir une valeur précise pour l'entropie. Il est intéressant de voir que, malgré cette imprécision, l'algorithme d'échange converge assez rapidement (figure 3) et que les plans ainsi construits ont les propriétés attendues, à savoir un remplissage uniforme de l'espace des paramètres, comme nous allons le voir dans le paragraphe 5.

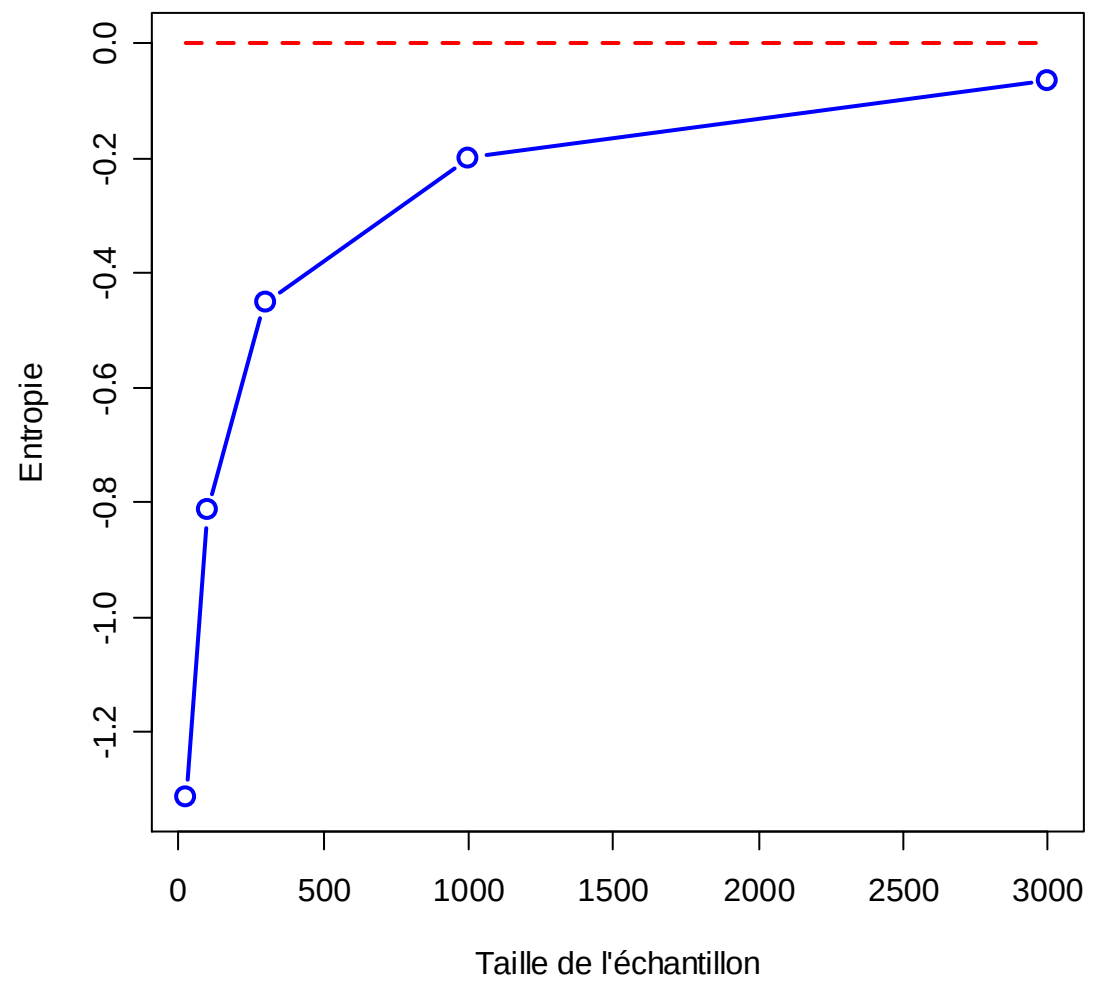

**Fig. 2. Evolution de l'entropie (MCGauss) en fonction de la taille de l'échantillon pour des tirages aléatoires en dimension 3**

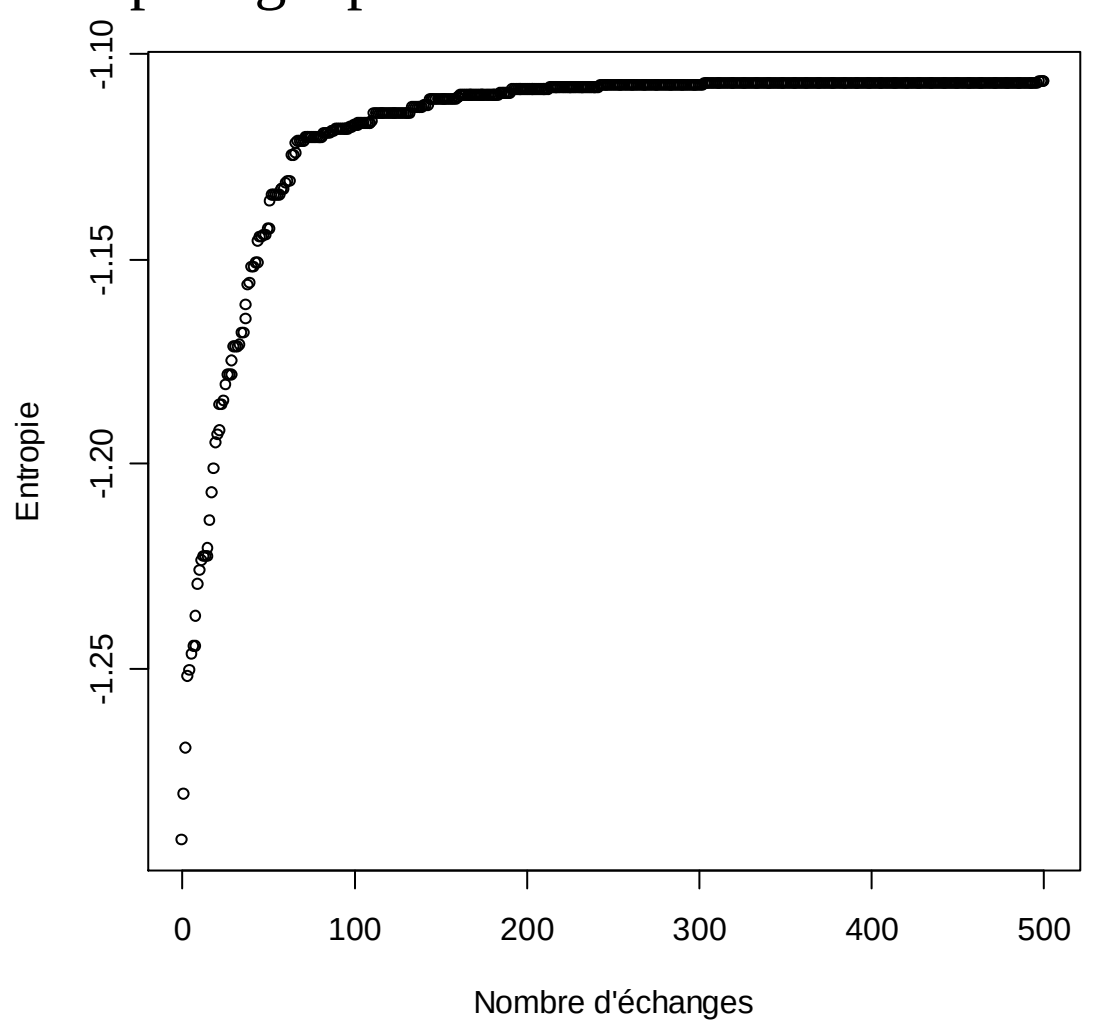

**Fig. 3. Evolution de l'entropie (MCGauss) au cours de l'algorithme d'échange pour des plans de taille 30 en dimension 3**

# 3. Estimation de l'entropie par plus proche voisin

Lorsque la dimension d augmente, la méthode d'estimation par Monte Carlo devient moins performante lorsque l'échantillon est de petite taille, comme cela est le cas pour un plan d'expériences numériques. Etant donnée que la densité doit être ré-estimée à chaque échange, cette méthode requière un temps de calcul prohibitif. Kozachenko et Leonenko (1987) proposent une estimation de l'entropie directement basée sur la répartition des points sans passer par l'estimation de la fonction de densité,

$$\hat{H}(f) = \frac{d}{n}\sum_{i=1}^{n}\ln \rho_i + \ln\left[\frac{\pi^{d/2}}{\Gamma(d/2+1)}\right] + C_E + \ln(n-1)$$

où $C_E \approx 0.5772$ est la constante d'Euler, $\Gamma$ est la fonction Gamma et $\rho_i$ est la distance euclidienne entre $X_i$ et son plus proche voisin dans l'échantillon,

$$\rho_i = \min_{j \neq i, 1 \leq j \leq n} \left\| X_i - X_j \right\|$$

Ils montrent que cet estimateur est asymptotiquement sans biais et converge en moyenne quadratique pour des conditions faibles sur f. On constate effectivement sur la figure 4. Cependant les valeurs sont positives alors qu'elles devraient être négatives en théorie. Ce décalage doit certainement retranscrire le biais de l'estimation. L'algorithme d'échange (figure 5) est lui aussi convergent mais semble plus lent que pour la méthode précédente. Le temps de calcul gagné sur l'estimation de l'entropie se trouve en partie perdu pas un nombre d'échanges plus important.

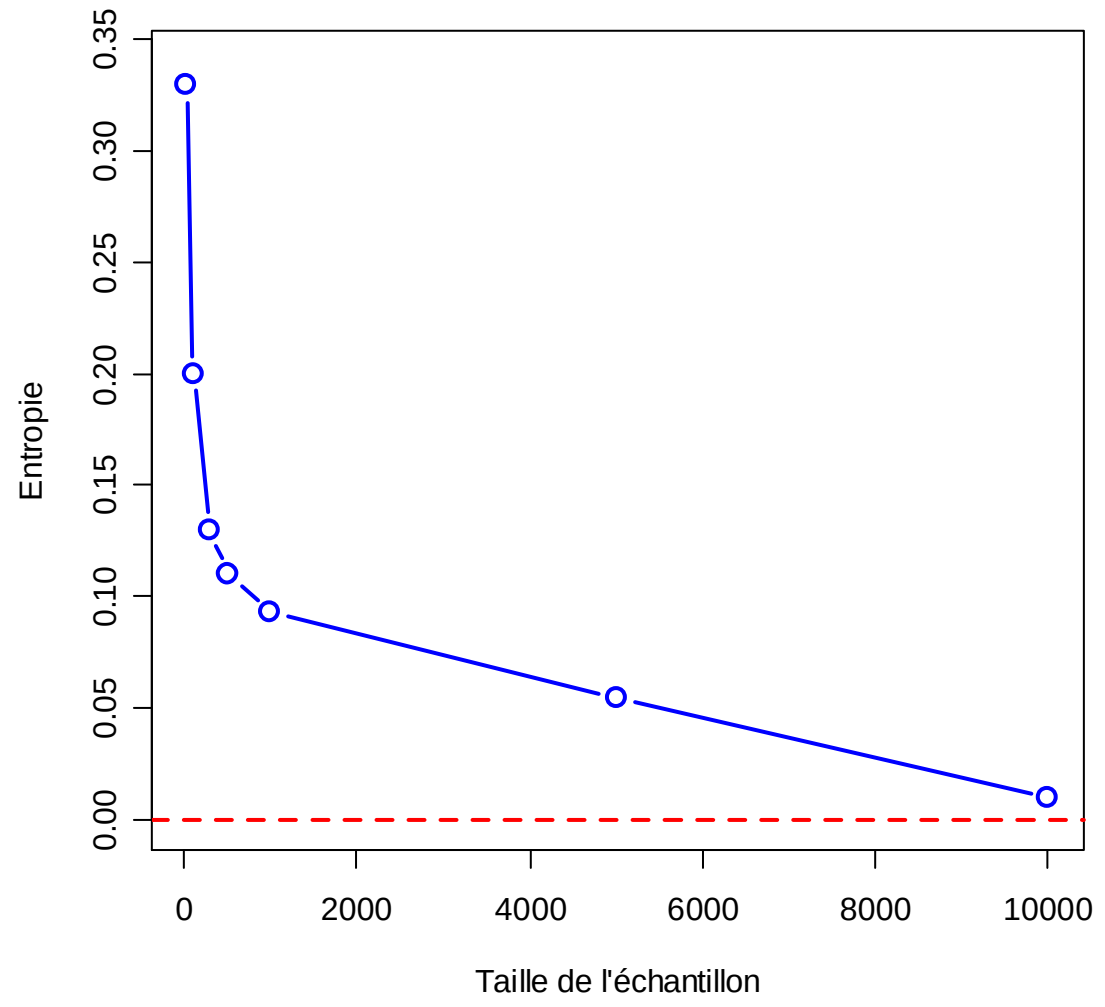

**Fig. 4. Evolution de l'entropie (PPV) en fonction de la taille de l'échantillon pour des tirages aléatoires en dimension 3**

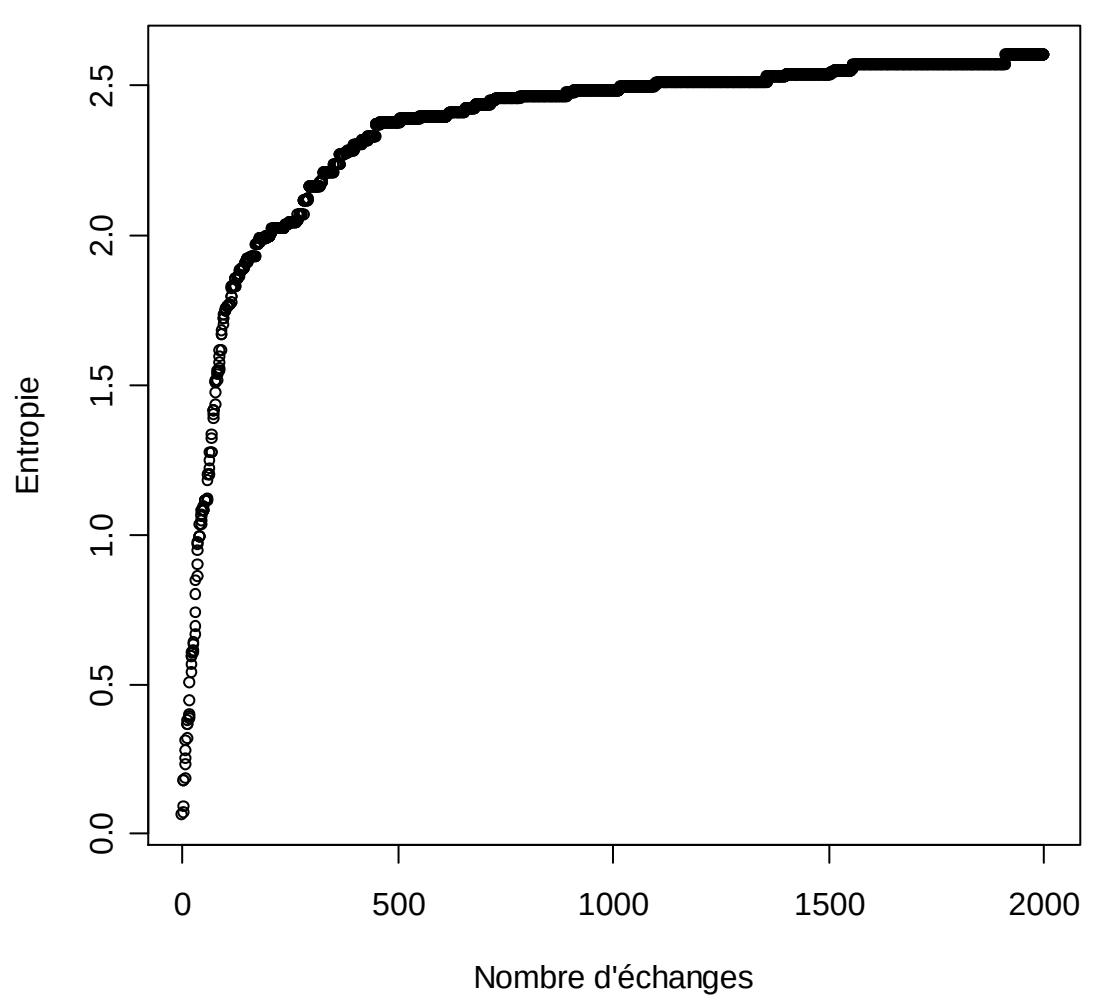

**Fig. 5. Evolution de l'entropie (PPV) au cours de l'algorithme d'échange pour des plans de taille30 en dimension 3**

Remarque : Une généralisation de cette estimation au $k^{ème}$ plus proche voisin a été proposée par Singh *et al.* (2003). Elle permet d'atténuer les grandes fluctuations de l'entropie engendrées les petites fluctuations des petites valeurs de $\rho_i$. Cette méthode présente un intérêt lorsqu'on travaille sur des données réelles car alors les distances $\rho_i$ ne sont pas connues avec précision. Ce qui n'est pas le cas dans le cadre de ce travail où les données simulées fournissent des distances exactes. Il semble alors préférable de prendre k=1, ce qui permet de discriminer les petites distances entre les points.

# 4. Comparaison des plans

Afin de juger de l'intérêt de la méthode de construction, nous avons construit des plans en dimension 2, 3 et 10 pour une taille standard fixée à $n=10\times d$.

Le plan issu de l'algorithme d'échange dépend plus ou moins du plan initial choisi aléatoirement. Afin de réduire cette dépendance, plusieurs initialisations sont testées et le meilleur plan est sélectionné. Dans les paragraphes suivants, plusieurs plans sont construits mais à partir d'une seule initialisation de façon à étudier l'influence du plan initial.

## 4.1. Amélioration de la distribution initiale

L'objectif de la méthode de construction est un remplissage uniforme de l'espace des paramètres. On constate visuellement sur la figure 6 que cet objectif est atteint pour les deux méthodes, et ce, quelle que soit l'initialisation du plan. On remarque que les plans finaux présentent les mêmes caractéristiques. Les points du plan sont disposés au bord du domaine mais aussi à l'intérieur à la façon d'une grille régulière légèrement perturbée.

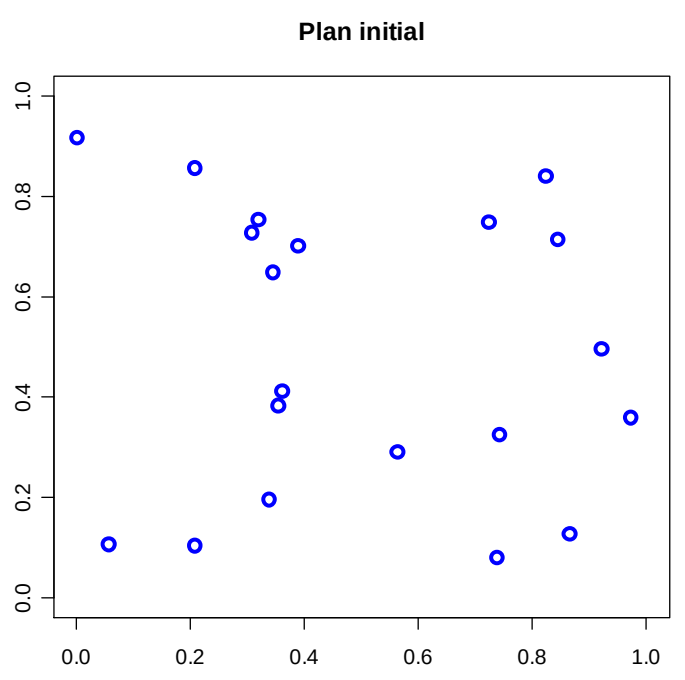

Initialisation aléatoire

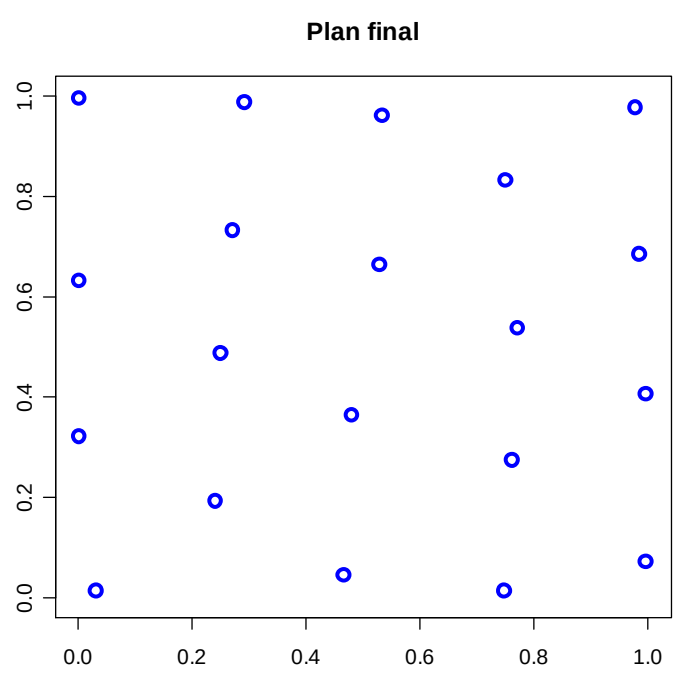

Méthode de Monte Carlo avec noyaux gaussiens

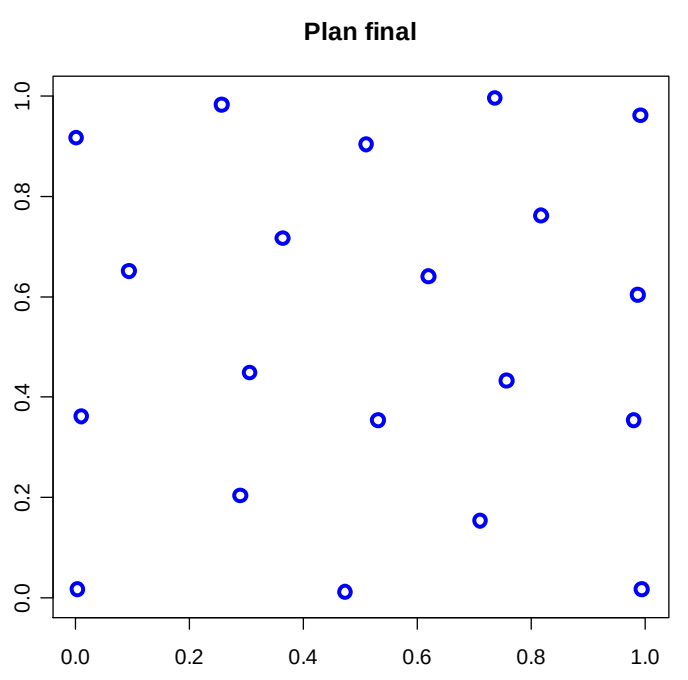

Méthode des plus proches voisins

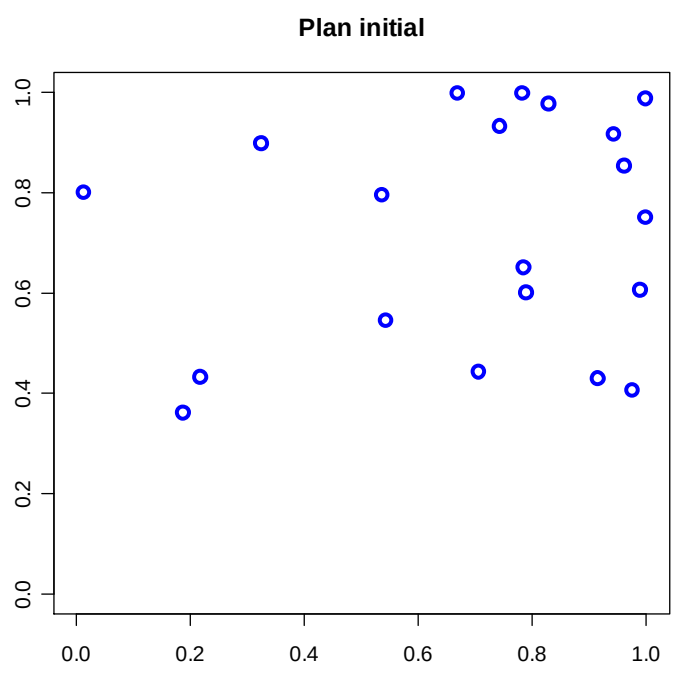

Initialisation suivant une loi Bêta

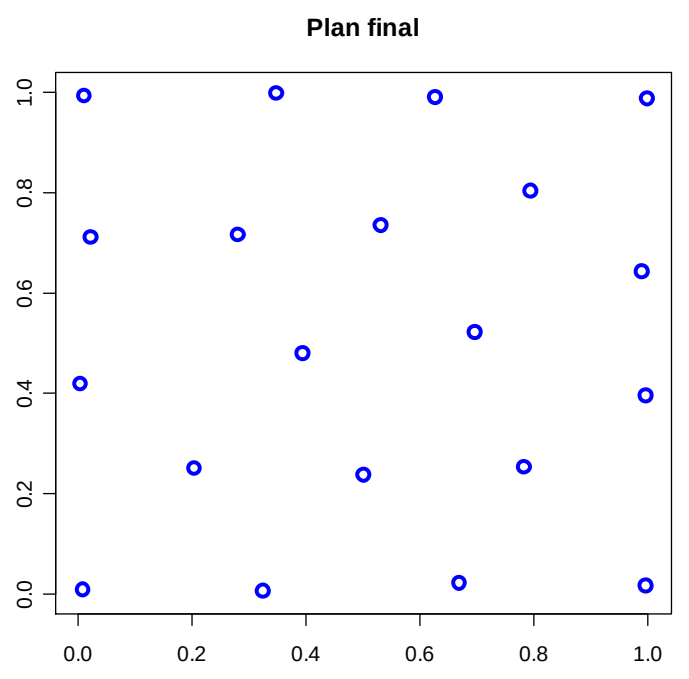

Méthode de Monte Carlo avec noyaux gaussiens

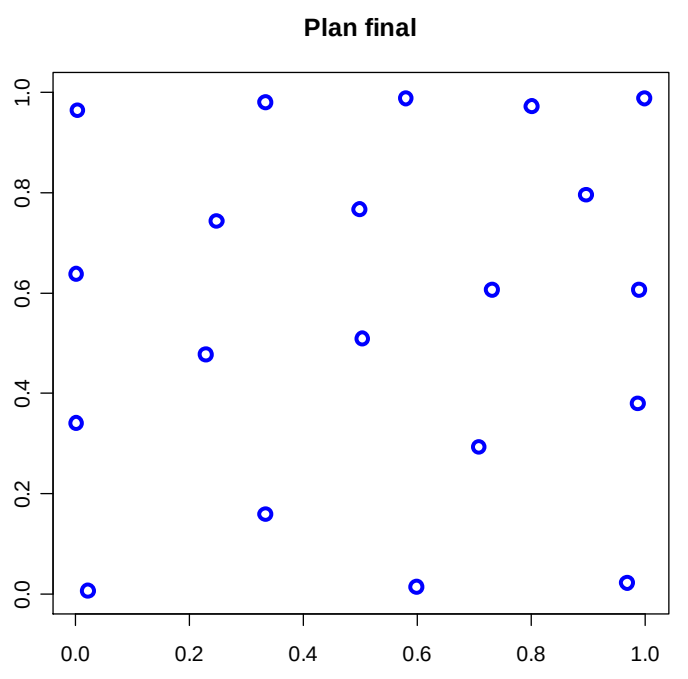

Méthode des plus proches voisins

**Fig. 6. Résultat de l'algorithme d'échange pour un plan de dimension 2 et de taille 20**

Chaque paramètre est testé sur un grand nombre de niveaux avec toutefois des répétitions aux extrémités. De même que le critère maximin (Koelher & Owen, 1996), le critère proposé ici privilégie les bords du domaine et non l'intérieur (cf. figure 7).

| Projection d'un plan 10d sur chaque axe | Projection d'un plan 10d sur le plan X1X2 |
|---|---|
| 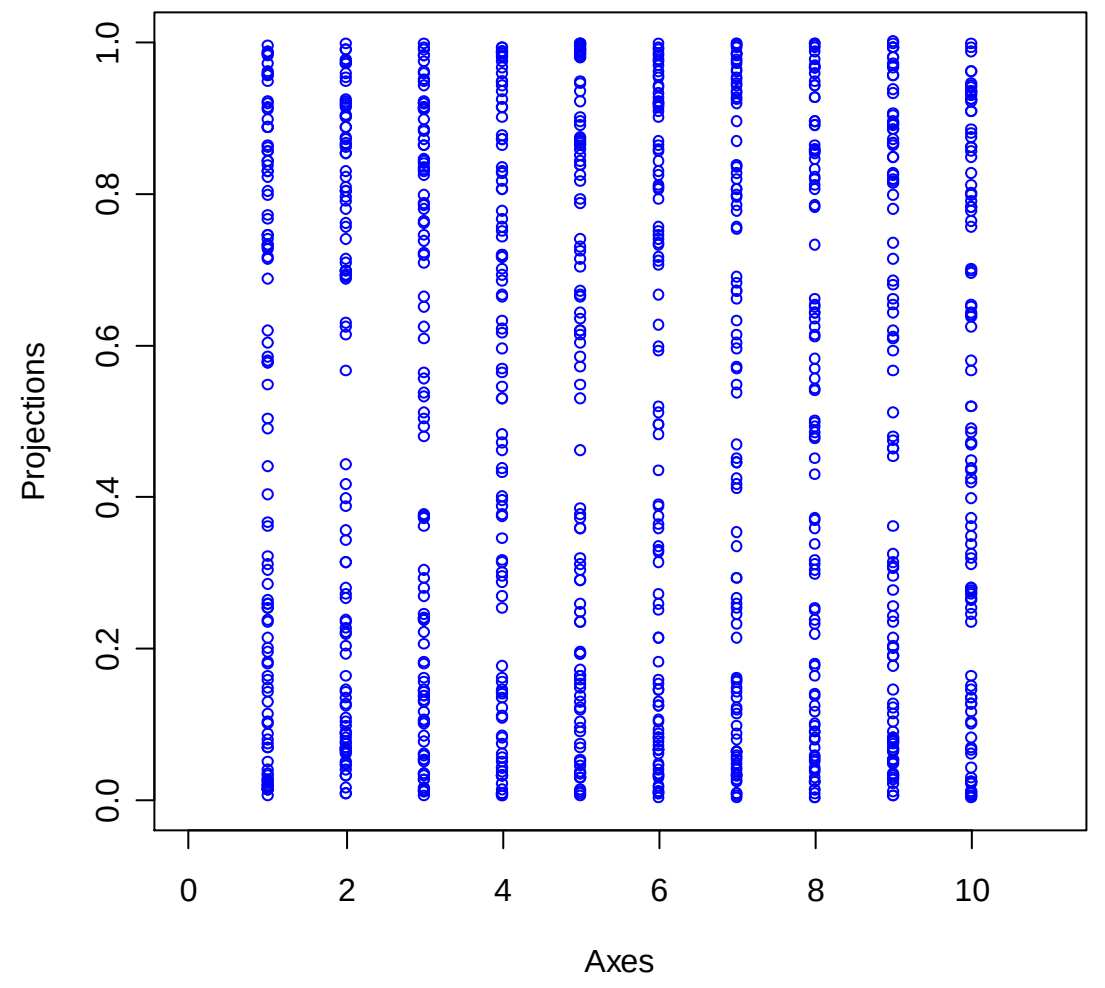 | 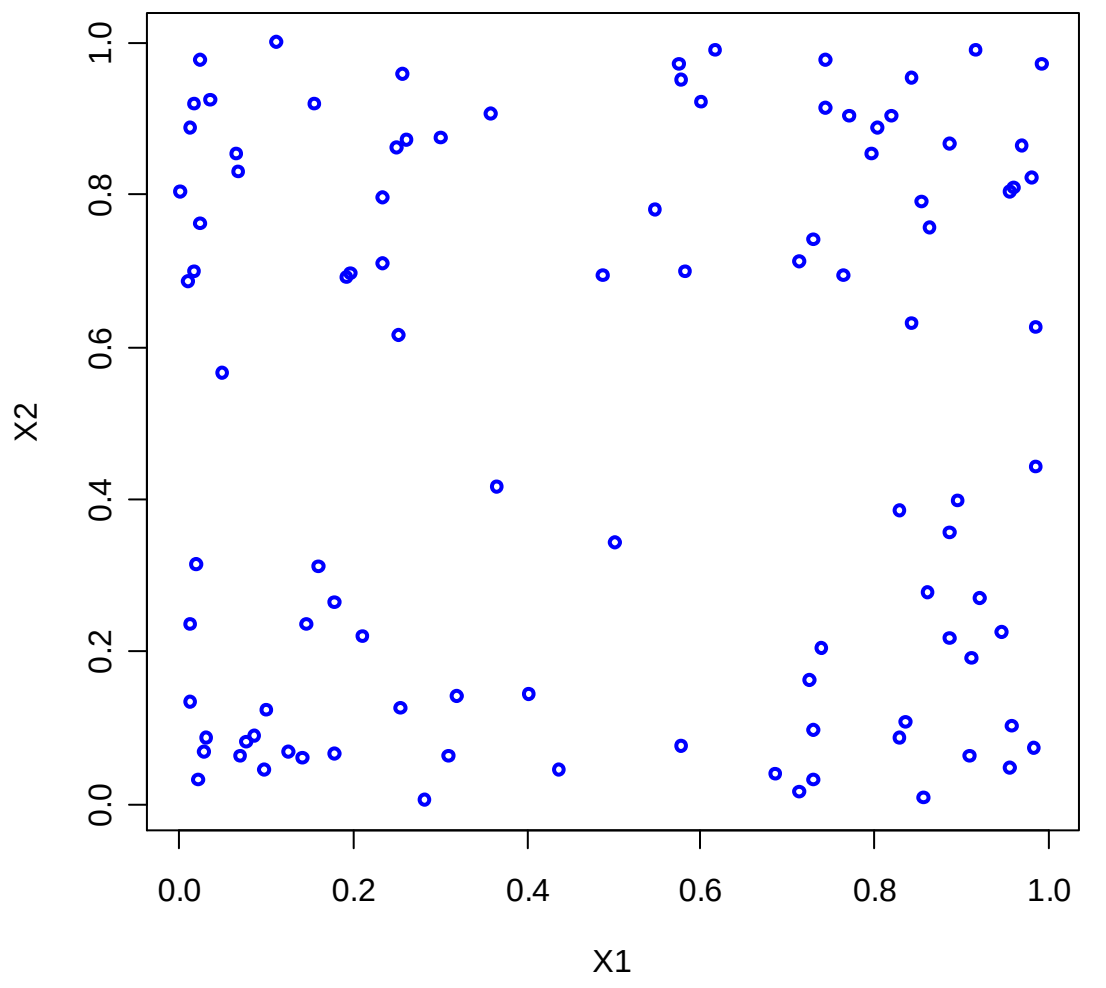 |

Plan construit par la méthode Monte Carlo avec noyaux gaussiens

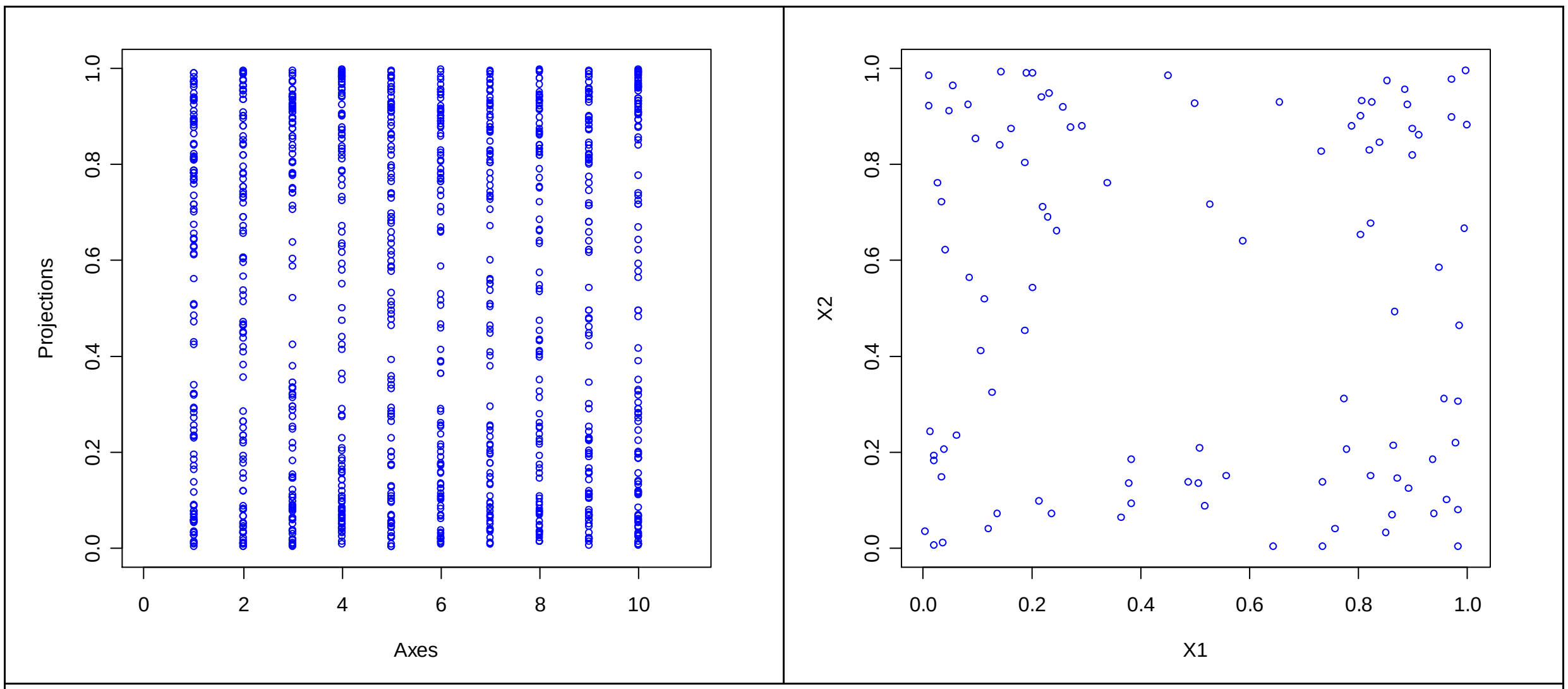

Plan construit par la méthode des plus proches voisins

**Fig. 7. Projections de plans de dimension 10 et de taille 100**

### 4.2. Critères et plans usuels

**Les critères**

Etudier l'uniformité d'une distribution de points lorsque d≥2, ne peut se faire visuellement. Il est donc utile de se référer à des critères afin de pouvoir décider si une distribution est uniforme ou non ou/et si elle vérifie également le bon remplissage de l'espace. Il existe pour cela des critères *intrinsèques* aux plans d'expériences − c'est-à-dire des critères qui ne préjugent en rien de la qualité de la surface de réponse déterminée par la suite à partir de ces plans − qui se prêtent particulièrement bien à cet objectif. Nous avons choisi de tester uniquement les critères d'uniformité les plus courants.

- La mesure de recouvrement (Cov). Elle permet de mesurer l'écart entre les points du plan et ceux d'une grille régulière (Gunzburger *et al.*, 2004). Ce critère est nul pour une grille régulière. L'objectif est donc de le minimiser pour se rapprocher d'une grille régulière, et ainsi assurer le remplissage de l'espace, sans toutefois l'atteindre pour respecter une distribution uniforme.
- La distance maximin (Mindist). Johnson *et al.* (1990) ont introduit les distances maximin et minimax afin de construire des plans répondant à la question de remplissage de l'espace. Le critère maximin consiste à maximiser la distance minimale entre deux points du plan.
- La discrépance (DL2, DC2). Contrairement aux deux critères précédents, la discrépance n'est pas basée sur la distance entre les points. Elle permet de mesurer l'écart entre la fonction de répartition empirique des points du plan et celle de la loi uniforme. Il existe différentes mesures de discrépance (Niederreiter, 1987, Thiémard, 2000). Nous retenons la discrépance en norme L2 et la discrépance en norme L2 centrée.

**Les plans usuels**

Les plans construits ici sont comparés avec les plans usuellement utilisés en expériences numériques. Nous conseillons la lecture des travaux de Koehler & Owen (1996) et Franco (2008) pour un état de l'art concernant les « space filling designs ». On notera que certains de ces plans sont optimaux pour les critères cités ci-dessus.

- Aléatoire : plans construits par tirage aléatoire dans le cube unité.
- LH : plans construits à partir d'hypercubes latins (sans critère d'optimisation) (Stein, 1987).
- Discrépance : suites à faible discrépance (Sobol, Niederreiter, Hammersley, Halton) (Niederreiter, 1987, Thiémard, 2000).
- Dmax : plans à entropie maximale (Shewry & Wynn, 1987, Currin *et al.*, 1988) construits de façon à maximiser le déterminant d'une matrice de covariance. Ces plans sont ainsi très utilisés lorsque la surface de réponse est ajustée par krigeage. Ils supposent cependant un modèle sous-jacent.
- Strauss : plans élaborés à partir d'un processus de Strauss qui considère de la répulsion entre les points de manière à remplir au mieux l'espace des paramètres (Franco, 2008)
- Maximin : plans optimaux pour le critère maximin (Johnson *et al.*, 1990).
- MCGauss : plans d'information KL minimale construits par la méthode décrite précédemment
- PPV : plans construits par estimation de l'entropie par plus proche voisin

En dimension 3 (figure 8), les plans d'information KL minimale donnent les meilleures valeurs exceptées pour le critère de discrépance. Ce résultat est inattendu, car les critères d'entropie et de discrépance visent tous les deux à minimiser l'écart entre la distribution empirique et la loi uniforme alors que les critères de recouvrement et de distance maximin sont basés sur la distance entre les points. Ce résultat peut s'expliquer par la difficulté à évaluer la discrépance. En effet, les tests effectués avec différentes méthodes de calcul de la discrépance (centrée, symétrique, ...), ainsi que les résultats en dimension 10, donnent des résultats paradoxaux.
Bien souvent, les space filling designs perdent de leurs qualités en grande dimension. Les tests en dimension 10 (figure 9) montrent que cela n'est pas le cas des plans d'information KL minimale. Ils semblent concurrencer les plans maximin traditionnellement utilisés en phase exploratoire, et ceci même pour le critère de distance maximin.

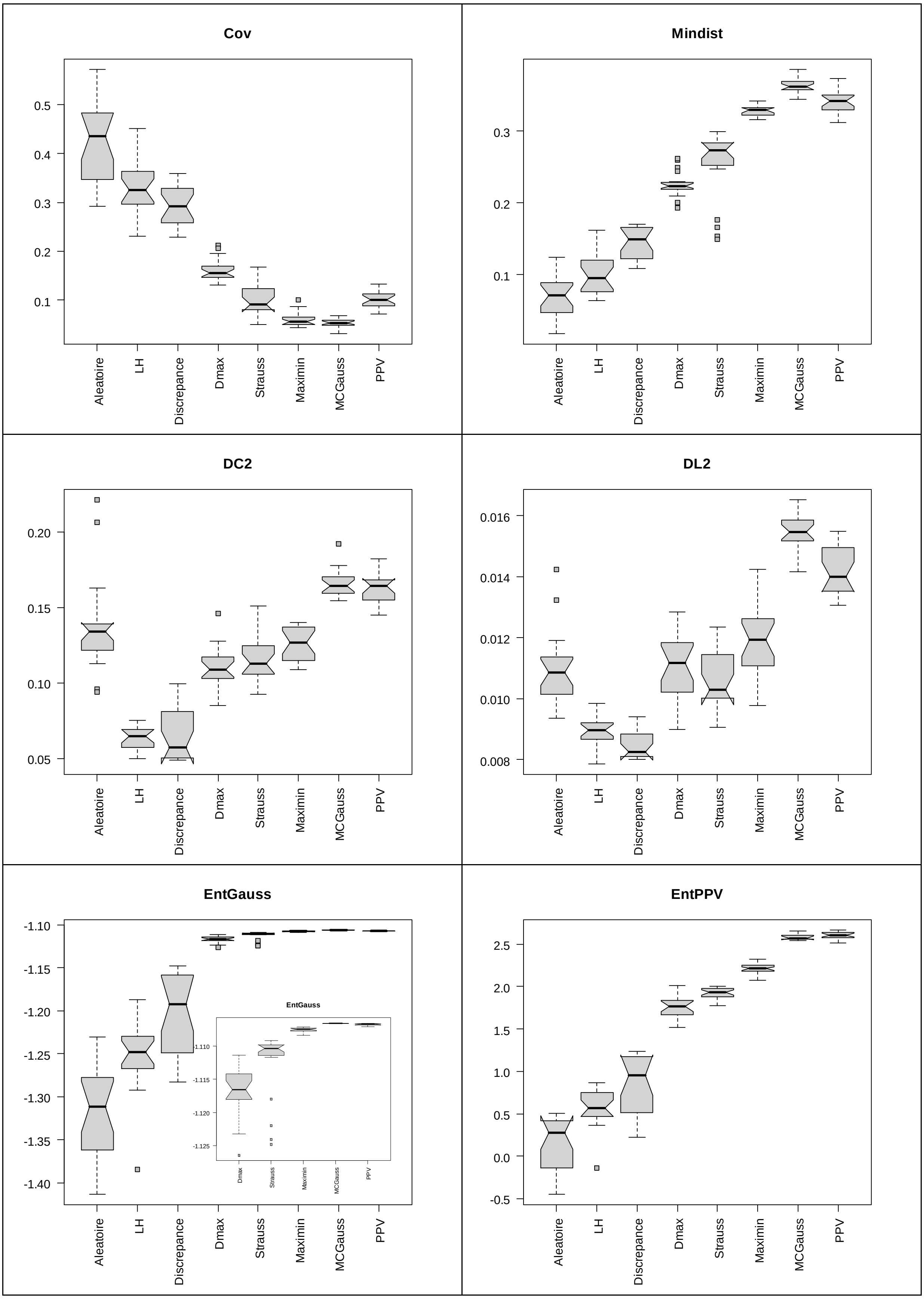

**Fig. 8. Représentation des critères usuels calculés sur 20 plans à 30 points en dimension 3**

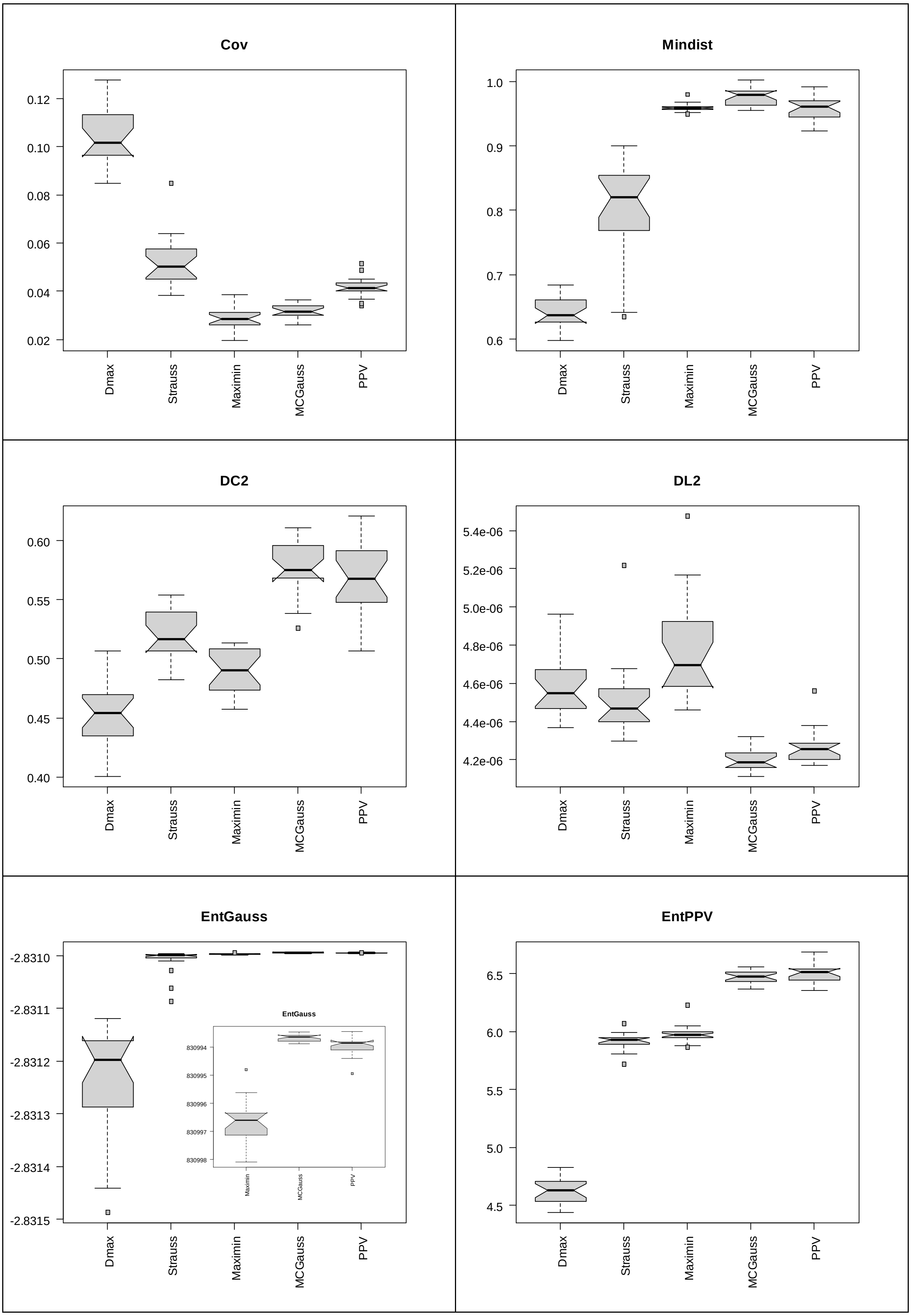

**Fig. 9. Représentation des critères usuels calculés sur 20 plans à 100 points en dimension 10**

Les plans d'information KL minimale sont incontestablement les meilleurs au regard des critères usuels. Les plans construits à l'aide de la méthode de Monte Carlo avec noyaux gaussiens sont légèrement plus performants que ceux construits par plus proches voisins.

### 4.3. Critère ALM

Le critère ALM (Arbres de Longueur Minimale) permet, non pas de mesurer la répartition des points d'un plan dans un espace multidimensionnel, mais de la qualifier et ainsi classer les plans selon leur structure : grille, aléatoire, en amas, … , suivant la cartographie empirique de la figure 10. Ce critère a été proposé par Franco *et al.* (2008) suite aux travaux de Wallet *et al.* (1998).

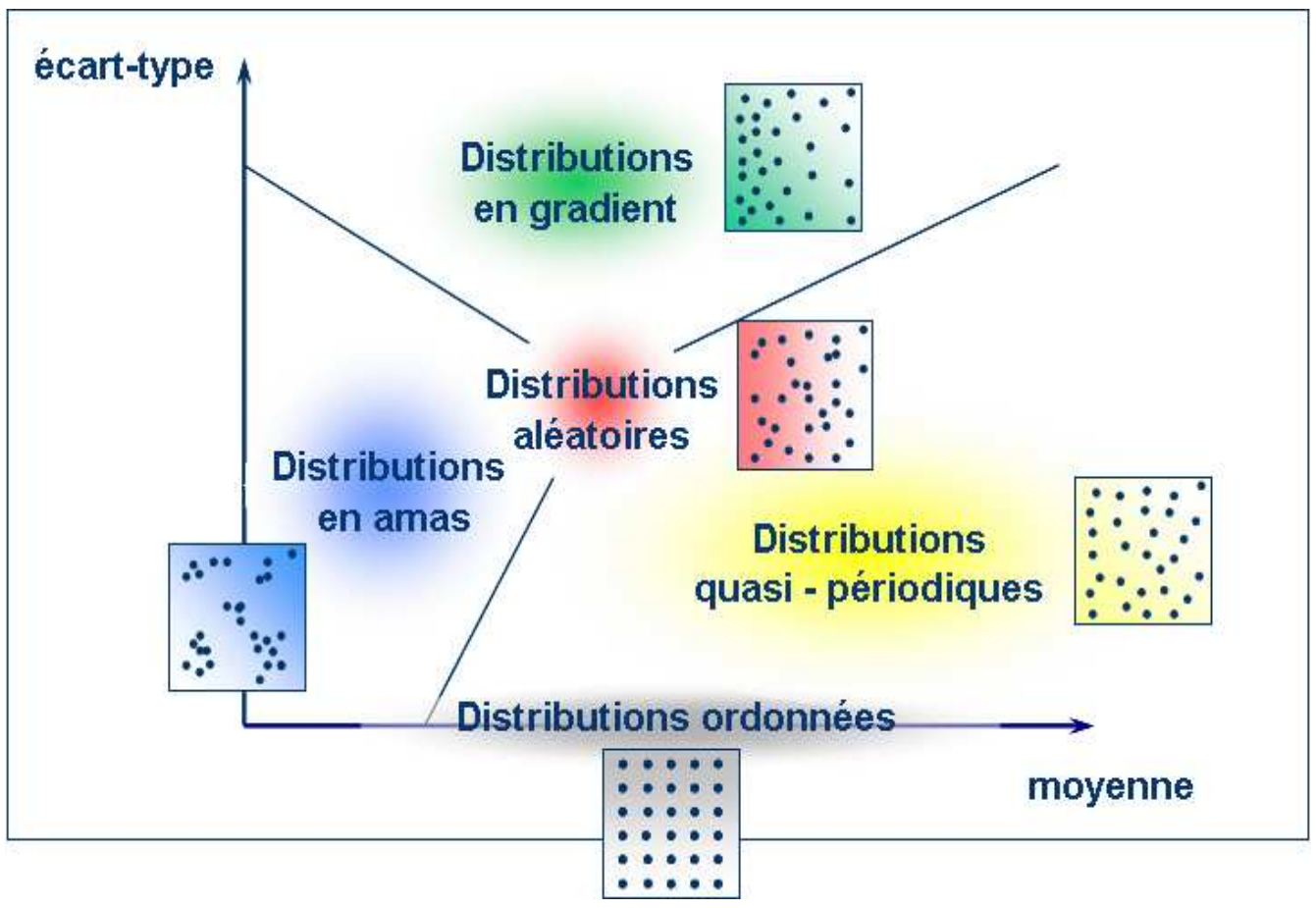

**Fig. 10. Cartographie empirique construite à partir du critère d'ALM**

La cartographie dépend de la moyenne et l'écart-type des longueurs des arêtes des ALM associés aux plans étudiés. Les plans de la zone de quasi-périodicité présentent le meilleur compromis entre grille régulière (remplissage de l'espace) et distribution aléatoire (uniformité). Les plans d'information KL minimale fournissent ici aussi les meilleurs résultats (figures 11 et 12) avec une nette démarcation en dimension 3. Ce résultat confirme les remarques sur les caractéristiques des plans de dimension 2 (§5.1). Les plans construits à l'aide de la méthode de Monte Carlo avec noyaux gaussiens sont là encore, plus performants que ceux construits par plus proches voisins.

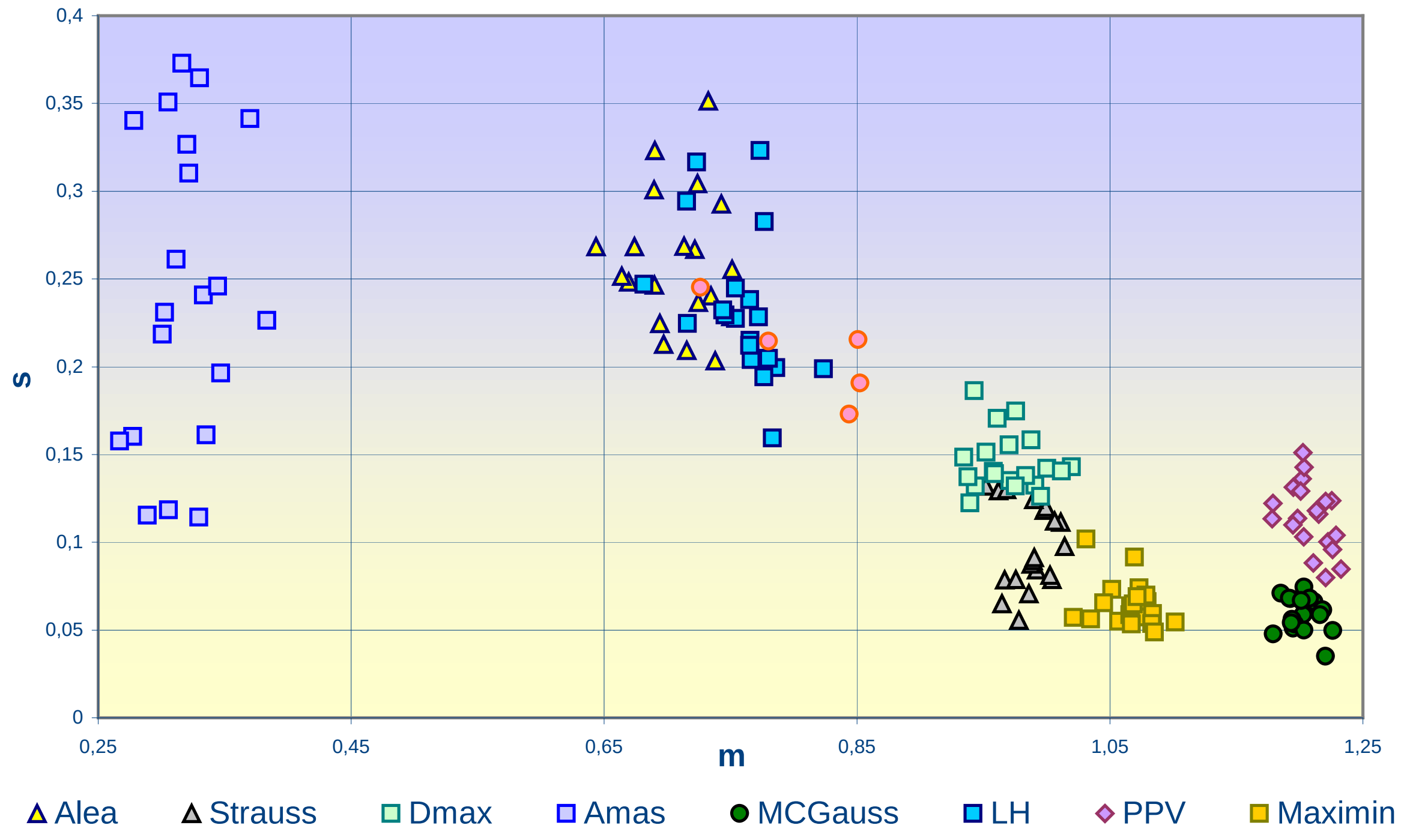

**Fig. 11. Représentation du critère ALM calculé sur 20 plans à 30 points en dimension 3**

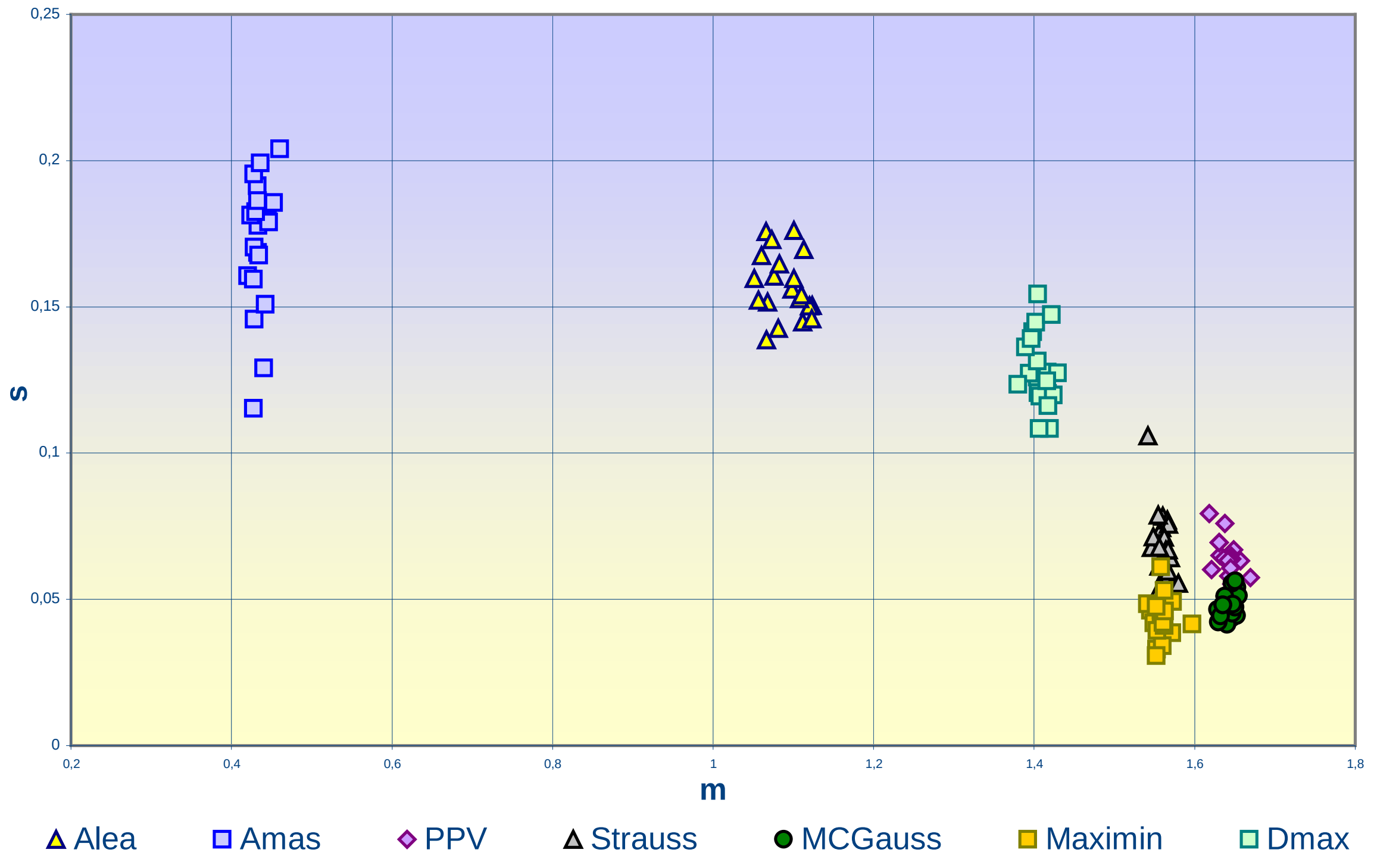

**Fig. 12. Représentation du critère ALM calculé sur 20 plans à 100 points en dimension 10**

# 5. Conclusions

L'objectif du critère présenté dans ce rapport est de minimiser l'écart entre la distribution empirique des points du plan et la loi uniforme. Minimiser l'information de Kullback-Leibler pour optimiser ce critère revient à maximiser l'entropie. Ainsi, toute la technique de construction des plans repose sur l'estimation de l'entropie. Celle-ci se fait par deux méthodes différentes. La première est une méthode de Monte Carlo dans laquelle la fonction de densité est estimée à l'aide de noyaux gaussiens. La deuxième évite l'estimation de la fonction de densité en utilisant une méthode basée sur la distance aux plus proches voisins. Les tests effectués montrent que les plans d'information KL minimale répondent à l'objectif principal, à savoir une bonne représentation du domaine de simulation par les points du plan, et ce, même en grande dimension. Quel que soit le critère choisi, ces plans concurrencent les space filling desgins usuels et notamment les plans maximin habituellement utilisés en phase exploratoire.

Malgré des conditions éloignées de l'asymptotique, un critère basé sur l'estimation de l'entropie de Shannon fournit de très bons résultats, avec une légère avance pour la méthode de Monte Carlo. Cependant, lorsque la dimension augmente, la méthode d'estimation par Monte Carlo devient moins performante lorsque l'échantillon est de petite taille. De plus, étant donnée que la densité doit être ré-estimée à chaque échange, cette méthode requière un temps de calcul prohibitif. En revanche, Leonenko *et al.* soulignent que la méthode des plus proches voisins se prête particulièrement bien aux grandes dimensions.

Une perspective consiste à définir le critère à partir d'une entropie différente de celle de Shannon, par exemple, l'entropie de Rényi qui peut être estimée à partir des arbres de longueur minimal (Hero *et al.*, 2002), ou bien l'entropie de Tsallis qui, à l'ordre deux et dans le cas d'une méthode d'estimation de la densité par noyaux gaussiens, s'écrit de façon analytique comme une somme des noyaux et évite ainsi l'erreur d'approximation de la méthode de Monte Carlo (Bettinger *et al.*, 2008).

# REFERENCES

[1] Ahmad I.A., Lin P.E. (1976). A nonparametric estimation of the entropy for absolutely continous distributions. *IEEE Trans. Information Theory*, 35, pp. 688-692.

[2] Ahmed N.A, Gokhale D.V. (1989). Entropy expressions and their estimators for multivariate distributions. *IEEE Transactions on Information Theory*, 35(3), pp. 688-692.

[3] Beirlant J., Dudewicz E.J., Györfi L., Van Der Meulen E.C. (1997). Nonparametric entropy estimation : an overview. Int. *J. Math. Stat. Sci.*, 6(1) 17-39.

[4] Bettinger R., Duchêne P., Pronzato L., Thierry E. (2008). Design of experiments for response diversity. In *Proc. 6th International Conference on Inverse Problems in Engineering (ICIPE), Journal of Physics: Conference*

*Series}*, Dourdan (Paris), 15-19 juin 2008, to appear. http://hal.archives-ouvertes.fr/hal-00290418/fr/

[5] Chaloner K., Verdinelli I. (1995). Bayesian Experimental Design: A review. *Statist. Sci.*, 10, 237-304.

[6] Currin C., Mitchell T., Morris M., Ylvisaker D. (1988). A bayesian approach to the design and analysis of computer experiments. ORNL Technical Report 6498, available from the national technical information service, Springfield, Va. 22161.

[7] Dimitriev Y.G, Tarasenko F.P. (1973). On the estimation functions of the probability density and its derivatives. *Theory Probab. Appl.*, 18, 628-633.

[8] Dudewicz E.J., Van Der Meulen E.C. (1981). Entropy-Based tests of uniformity. J. *Amer. Statist. Assoc.*, 76, 967-974.

[9] Franco J (2008). *Planification d'expériences numériques en phase exploratoire pour des codes de calculs simulant des phénomènes complexes*. Thèse présentée à l'Ecole Nationale Supérieure des Mines de Saint-Etienne

[10] Franco J., Vasseur O., Corre B., Sergent M. (2008). Minimum Spanning Tree : A new approach to assess the quality of the design of computer experiments. To appear in *Chemometrics and Intelligent Laboratory Systems*.

[11] Gunzburger M., Burkardt J. (2004). Uniformity measures for point sample in hypercubes. https://people.scs.fsu.edu/~burkardt/pdf/ptmeas.pdf

[12] Harner, E. James, Singh H., Li, S., and Jun Tan (2004) Computational Challenges in Computing Nearest Neighbor Estimates of Entropy for Large Molecules. Proceeding of the 36th Symposium on the Interface: Computational Biology and Bioinformatics, may 26-29.

[13] Hero A., Bing Ma, Michel O., Gorman J. (2002). Applications of entropic spanning graphs. *Signal Processing Magazine, IEEE*, 19, 85 – 95.

[14] Joe H. (1989). Estimation of entropy and other functional of multivariate density. *Ann. Int. Statist. Math.*, 41, 683-697.

[15] Johnson M.E., Moore L.M., Ylvisaker D. (1990). Minimax and maximin distance design. *J. Statist. Plann. Inf.*, 26,131-148.

[16] Koehler J.R., Owen A.B (1996). *Computer Experiments*. Handbook of statistics, 13, 261-308.

[17] Kosachenko L.F., Leonenko N.N. (1987). Sample estimate of entropy of a random vector. *Problem of Information Transmission*, 23, 95-101.

[18] Kullback S., Leibler R.A. (1951). On information and sufficiency. *Ann. Math. Statist.*, 22 79-86.

[19] Leonenko N, Pronzato L, Savani V. A class of Rényi information estimators for multidimensional densities. *Annals of Statistics* , to appear.

[20] Niederreiter H. (1987). Point sets and sequences with small discrepancy. *Monasth. Math.*, 104, 273-337.

[21] Scott D.W. (1992). *Multivariate Density Estimation : Theory, practice and visualization*, John Wiley & Sons, New York, Chichester.

[22] Sebastiani P. & Wynn H.P. (2000), Maximum entropy sampling and optimal Bayesian experimental design. *J. Royal Statist. Soc.*, 62, 145-157

[23] Shewry M.C., Wynn H.P. (1987). Maximum Entropy Sampling. *J. Appl. Statist.*, 14, 165-170.

[24] Silverman B.W. (1986). *Density estimation for statistics and data analysis*. Chapman & Hall, London.

[25] Singh, H., Misra, N., Hnizdo, V., Fedorowicz, A. & Demchuk, E. (2003) Nearest neighbor estimates of entropy. *Am. J. Math. Manage. Sci.*, 23, pp. 301-321.

[26] Stein M. (1987). Large sample properties of simulations using latin hypercube sampling. *Techometrics*, 29, 143-151.

[27] Thiémard E. (2000). *Sur le calcul et la majoration de la discrépance à l'origine*. Thèse présentée au département de mathématiques de l'école polytechnique fédérale de Lausanne

[28] Wallet F., Dussert C. (1998). Comparison of spatial point patterns and processes characterization methods. *Europhysics Lett.*, 42, 493-498.

## ANNEXE 1 : Calcul de la probabilité d'un noyau non nul dans la méthode de Monte Carlo par noyaux uniforme ou d'Epanechnikov

Le vecteur z prend les valeurs suivantes

$$z = \frac{x - X_i}{h}$$

où $X_i$ est un point du plan donc tel que ses coordonnées $X_i^j$, j=1,...d suivent une loi uniforme sur [0,1] et x est le point où on cherche à estimer la fonction de densité. Le noyau est non nul si

$$\|z\| \leq 1 \Leftrightarrow \|z\|^2 \leq 1$$

Dans la méthode de Monte Carlo utilisée, la fonction de densité est estimé uniquement aux points du plan, ce qui signifie que les coordonnées $x^j$, j=1,...,d suivent aussi une loi uniforme sur [0,1]. Supposons que $x \neq X_i$ (sinon le noyaux est non nul et prend la valeur $\alpha$), alors

$$-1 \leq x^j - X_i^j \leq 1 \Rightarrow 0 \leq \left(x^j - X_i^j\right)^2 \leq 1 \Rightarrow 0 \leq \sum_{j=1}^{d} \left(x^j - X_i^j\right)^2 \leq d \Rightarrow 0 \leq \|z\|^2 \leq d/h^2$$

Donc la norme de z au carré suit une loi uniforme sur $[0,d/h^2]$, d'où

$$P\left(\|z\|^2 \leq 1\right) = h^2/d.$$

Or $h^2 = \frac{1}{12}\frac{1}{n^{2/(d+4)}}$ et $n = 10d$

donc

$$P\left(\|z\|^2 \leq 1\right) = \frac{1}{12d}\frac{1}{(10d)^{2/(d+4)}}$$

| d | 1 | 2 | 3 | 4 | 5 |
|---|---|---|---|---|---|
| $P\left(\|z\|^2 \leq 1\right)$ | 3,3E-02 | 1,5E-02 | 1,1E-02 | 8,3E-03 | 7,0E-03 |
| d | 6 | 7 | 8 | 9 | 10 |
| $P\left(\|z\|^2 \leq 1\right)$ | 6,1E-03 | 5,5E-03 | 5,0E-03 | 4,6E-03 | 4,3E-03 |

## ANNEXE 2 : Calcul de la constante de normalisation dans l'estimation de le densité par noyaux d'Epanechnikov

Prenons le cas où le noyau est produit de noyaux d'Epanechnikov, alors

$$\int_{R^d} \mathcal{K}(x_1,\dots,x_d) = \int_{R^d} \prod_{k=1}^{d} K(x_k) = \left(\int_R K(x)dx\right)^d = \left(\alpha\int_{-1}^{1} 1-\|x\|^2 dx\right)^d = \left(2\alpha\int_0^1 1-x^2 dx\right)^d = \left(\alpha\frac{4}{3}\right)^d$$

En posant l'intégrale égale à 1, on obtient $\alpha=3/4$ et ce quelle que soit la dimension.

Prenons le cas où le noyaux est sphérique de dimension 2, alors

$$\iint_{R^2} \mathcal{K}(x_1, x_2)dx_1dx_2 = \alpha\iint_{\{\|x\|\leq 1\}} \left(1-\|x\|^2\right)dx$$

On passe en coordonnées polaires

$$\begin{cases} x = r\cos\theta \\ y = r\sin\theta \end{cases} \Rightarrow J = \begin{pmatrix} \cos\theta & -r\sin\theta \\ \sin\theta & r\cos\theta \end{pmatrix} \Rightarrow \det J = r \text{ et } 1-\|x\|^2 = 1-r^2$$

d'où

$$\iint_{R^2} \mathcal{K}(x_1, x_2)dx_1dx_2 = \alpha\int_0^{2\pi}\int_0^1 r(1-r^2)drd\theta = \alpha\frac{\pi}{2}$$

Donc $\alpha=2/\pi$.

Prenons le cas où le noyaux est sphérique de dimension 3, alors

$$\iiint_{R^3} \mathcal{K}(x_1, x_2, x_3)dx_1dx_2dx_3 = \alpha\int\iint_{\{\|x\|\leq 1\}} \left(1-\|x\|^2\right)dx$$

On passe en coordonnées sphériques

$$\begin{cases} x = r\sin\theta\cos\varphi \\ y = r\sin\theta\sin\varphi \\ z = r\cos\theta \end{cases} \Rightarrow J = \begin{pmatrix} \sin\theta\cos\varphi & r\cos\theta\cos\varphi & -r\sin\theta\sin\varphi \\ \sin\theta\sin\varphi & r\cos\theta\sin\varphi & r\sin\theta\cos\varphi \\ \cos\theta & -r\sin\theta & 0 \end{pmatrix} \Rightarrow \det J = r^2\sin\theta$$

$$\iiint_{R^3} \mathcal{K}(x_1, x_2, x_3)dx_1dx_2dx_3 = \alpha\int_0^{2\pi}\int_0^{\pi}\int_0^1 (1-r^2)r^2\sin\theta drd\theta d\varphi = \alpha\frac{8\pi}{15}$$

Donc $\alpha=15/(8\pi)$.

Dans le cas où le noyaux est sphérique de dimension supérieure à 3, il faut faire une approximation numérique. Par exemple, la méthode de Monte Carlo donne une approximation de l'intégrale par

$$\frac{V_d}{N}\sum_{i=1}^{N} g(x_i)$$

où $g(x_i) = 1-\|x_i\|^2$, les $x_i$ sont N points tirer au hasard tels que $\|x_i\| \leq 1$ et $V_d$ est le volume de la boule unité en dimension d,

$$V_d = \frac{\pi^{d/2}}{(d/2)\Gamma(d/2)}$$

Si d est pair, d=2p, alors $V_d = \frac{\pi^p}{p\Gamma(p)} = \frac{\pi^p}{p(p-1)!} = \frac{\pi^p}{p!}$.

Si d est impair, d=2p+1, alors $V_d = \frac{\pi^{p+1/2}}{(p+1/2)\Gamma(p+1/2)} = \frac{\pi^{p+1/2}}{(p+1/2)} \frac{2^{2p}p!}{(2p)!\pi^{1/2}} = \pi^p \frac{2^{2p}p!}{(2p)!(p+1/2)}$

| **d** | 1 | 2 | 3 | 4 | 5 | 6 | 7 | 8 | 9 | 10 |
|---|---|---|---|---|---|---|---|---|---|---|
| **α** | 3/4 | 2/π | 15/(8π) | 0.61 | 0.67 | 0.77 | 0.95 | 1.23 | 1.66 | 2.38 |

**d pair**

```
d<-2
## volume de la boule unité
p<-d/2
V<-(pi^p)/factorial(p)

estimateur<-array(0,dim=c(1,10))
## Boucle sur estimateur
for (k in 1 :10) {
## echantillon
N<-100000
x<-array(runif(N*d,min=0,max=1),dim=c(d,N))
## Estimation
cpt<-0
for (i in 1 :N) {
    norme<-0
    for (j in 1 :d) {
    norme<-norme+x[j,i]^2
    }
    if (norme<=1) {
      estimateur[k]<-estimateur[k]+1-norme
      cpt<-cpt+1
    }
}
estimateur[k]<-estimateur[k]*V/cpt
}
estim<-mean(estimateur)
alpha <- 1/estim
```

**d impair**

```
d<-3
## volume de la boule unité
p<-floor(d/2)
V<-
(pi^p)*factorial(p)*(2^(2*p))/(factorial(2*p)*(p+0.5))

estimateur<-array(0,dim=c(1,10))
## Boucle sur estimateur
for (k in 1 :10) {
## echantillon
N<-100000
x<-array(runif(N*d,min=0,max=1),dim=c(d,N))
## Estimation
cpt<-0
for (i in 1 :N) {
    norme<-0
    for (j in 1 :d) {
    norme<-norme+x[j,i]^2
    }
    if (norme<=1) {
      estimateur[k]<-estimateur[k]+1-norme
      cpt<-cpt+1
    }
}
estimateur[k]<-estimateur[k]*V/cpt
}
estim<-mean(estimateur)
alpha <- 1/estim
```

Programme R